# Design and experimental study of superconducting left-handed transmission lines with tunable dispersion

E A Ovchinnikova[1], S Butz[2], P Jung[2], V P Koshelets[1,3], L V Filippenko[1,3], A S Averkin[1], S V Shitov[1,3], and A V Ustinov[1,2]

[1]National University of Science and Technology (MISIS), Leninsky prosp. 4, Moscow 119049, Russia
[2]Physikalisches Institut, Karlsruhe Institute of Technology (KIT), 76131 Karlsruhe, Germany
[3]Kotel'nikov Institute of Radio Engineering and Electronics, Moscow 125009, Russia
E-mail: elhed@yandex.ru

**Abstract.** We study microwave properties of a superconducting tunable coplanar waveguide (CPW). Pairs of Josephson junctions are forming superconducting quantum interference devices (SQUIDs), which shunt the central conductor of the CPW. The Josephson inductance of the SQUIDs is varied in the range of 0.08-0.5 nH by applying an external dc magnetic field. The central conductor of the CPW contains Josephson junctions connected in series that provide extra inductances; the magnetic field controlling the SQUIDs is weak enough not to influence the inductance of the chain of the single junctions. The circuit is designed to have left- and right-handed transmission properties separated by a variable rejection band; the band edges can be tuned by the magnetic field. We present transmission measurements on CPWs based on up to 120 Nb-$AlO_x$-Nb Josephson junctions. At zero magnetic field, we observed no rejection band in the frequency range of 8-11 GHz. When applying the magnetic field, a rejection band between 7 GHz and 9 GHz appears. The experimental data are compared with numerical simulations.

## 1. Introduction

Left-handed electromagnetic materials were considered in 1967 by Veselago [1]. Using Maxwell equations, it was shown that such materials exhibit a negative index of refraction. The medium was called "left" or "left-handed" (LH) because the electric field E, magnetic field H and the wave vector k form a left-handed system, while in ordinary media they form a right-handed system. According to Maxwell equations, the phase velocity of the LH media is opposite to both the direction of energy flow and the group velocity [2]. Since their first experimental realization in 2000 [3], the interest in left-handed metamaterials is unbroken and growing. One relatively simple way of implementing a left-handed medium is building of a planar transmission line. The electric permittivity ε and magnetic permeability μ of such a transmission medium can be modeled using distributed *L-C* networks [4]. The traditional two-wire model of a right-handed transmission line (RHTL) with inductance *L* on the central conductor and capacitance *C* on the shunt describes an electromagnetic medium with positive index of refraction, where *L* and *C* define positive equivalents for permeability and permittivity, respectively. A left-handed transmission line (LHTL) with negative permeability and permittivity can be created by simply exchanging the positions of *L* and *C* in the line [5].

## 2. Design and computer simulation

A transmission line can be described by the purely left-handed or right-handed two-wire model. However, in reality, it will always be of mixed type while LHTL or RHTL behaviour can dominate within a fixed frequency range [6]. In this work, we propose, design and test a transmission line, which is tunable from RHTL to LHTL under application of a dc magnetic field. The tunability of LHTLs has already been considered in some works, for example in Refs. [7] and [8]. In Ref. [7] the pass-band of the line is changed by applying a dc bias current, which tunes the kinetic inductance of a superconducting wire. The tunability has been also provided via applying a static electric field to thick films of barium-stronium-titanate (BST) which changes the electrical properties of the structure.

The suggested transmission line contains dc SQUIDs as tunable elements, which are known for their high sensitive to magnetic fields. The tuning of the dispersion is obtained via changing the Josephson inductance of the SQUIDs. Previously, Josephson junction arrays with two-junction unit-cells have been studied theoretically in Ref. [9]. Here, we focus on experimental study of a tunable rejection band in the transmission spectrum of the planar CPW. Our approach relies on tunability of SQUIDs, which has been recently experimentally demonstrated in Ref. [10].

The Josephson inductance of the dc SQUID is given by the relation from Ref. [11]:

$$L_j = \frac{\Phi_0}{2\pi I_c \cos\varphi} , \qquad (1)$$

here $\Phi_0$ is the magnetic flux quantum, $I_c$ is the maximum (zero-field) critical current of the junction and $\varphi$ is the superconducting phase difference. The phase difference is defining the superconducting current through the junction in a non-zero magnetic field. It is essential to notice that influence of the magnetic field on the inductance $L_{j1}$ of the series Josephson junctions in the main line is negligible because the applied magnetic field is normal to the plane of the electrodes of the junction. The magnetic flux coupled to the series junctions is negligibly small, therefore the inductance $L_{j1}$ of the junctions can be assumed as a constant for the whole range of magnetic fields used in this experiment.

The lumped-element equivalent circuit of a section of our transmission line is shown in Fig. 1. Capacitance $C_1$ and inductance $L_{j2}$ define the left-handed properties of the transmission line while capacitance $C_2$ and inductance $L_{j1}$ contribute to the right-handed behaviour. Since both chains $C_1$-$L_{j1}$ and $C_2$-$L_{j2}$ are resonators, one would expect the line may have, depending on exact parameters of $C's$ and $L's$, properties of either type in different frequency bands.

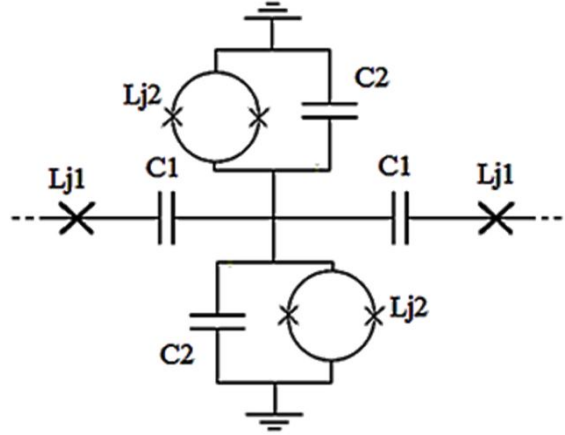

**Figure 1**. Scheme of unit cell. Crosses indicate Josephson junctions, circles are SQUID loops and $L_j$ denote the effective inductance of either SQUIDs or Josephson junctions. The experimental devices contains up to 20 cells in series that is long enough to treat the structure as a fraction of a transmission line within frequency range 4-12 GHz.

In order to understand the LHTR properties, one has to consider the dispersion relation f($\beta$) of travelling waves as in Ref. [7]. The propagation constant $\beta$ shows the phase and amplitude of electromagnetic wave along the propagation direction. An example of the frequency dependence of the propagation constant $\beta$ is shown in Fig. 2 that is calculated for the following parameters specific to our circuit: $C_1$= 2.1 pF, $C_2$=1.7 pF, $L_{j1}$= 80 pH. The tunable inductance $L_{j2}$ is controlled by a small uniform dc magnetic field, which induces dissipation-free persistent currents in the loops of the SQUIDs. Fig. 2a shows an example of the dispersion curve for $L_{j2}$ = 300 pH, while in Fig. 2b we use $L_{j2}$ = 190 pH.

One can see from Fig. 2 that the phase and group velocities, $V_{Ph}$ and $V_g$, have opposite signs in the lower part of the dispersion spectrum. This mean that the lower transmission band is the left-handed band. In contrast, phase and group velocities have the same direction above the upper cut-off frequency that mean the right-handed band. The lower and upper cut-off frequencies shown in the text below, are respectively the upper frequency of left-handed band and the lower frequency of the right-handed band. These cut-off frequencies can be expressed via the following equations from Ref. [12]:

$$f_{c1} = \frac{1}{2\pi\sqrt{L_{j1}C_1}} \ ; \qquad f_{c2} = \frac{1}{2\pi\sqrt{\frac{1}{2}L_{j2}C_2}} \ . \qquad (2)$$

It can be seen from Eq. (1) and Eq. (2) that changing $L_{j2}$ one can tune the upper frequency of the rejection band.

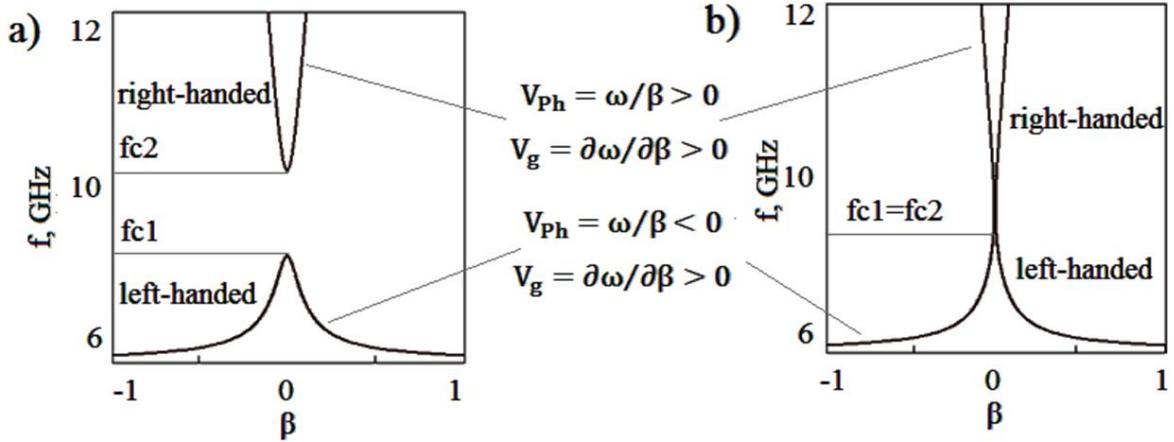

**Figure 2.** Calculated dispersion curves of LHTL. a) Josephson inductance $L_{j2}$= 300 pH leads to $f_{c1} \neq f_{c2}$; a rejection band occurs between frequencies $f_{c1}$ and $f_{c2}$; b) Josephson inductance $L_{j2}$= 190 pH results in $f_{c1} = f_{c2}$; the rejection band is absent.

We have started our study with numerical simulation of the transmission through an array of 20 unit cells from Fig. 1. The results of the simulation are shown in Fig. 3 for different values of $\varphi$, corresponding to different values of the magnetic field. One can see that the dip in the transmission between the left-handed (lower band) and the right-handed (higher band) indeed depends on the inductance $L_{j2}$, which is controlled via application of magnetic field, as expected from Eq. (1) and Eq. (2). In Fig. 3a only one, the left-handed pass-band is present. The application of magnetic field leads to the merge of the left and right-handed pass-bands as shown in Fig. 3b. At even larger field $\cos\varphi$ gets close enough to zero, the left-handed and right-handed bands of transmission are again separated by a rejection band. This case is depicted in Fig. 3c.

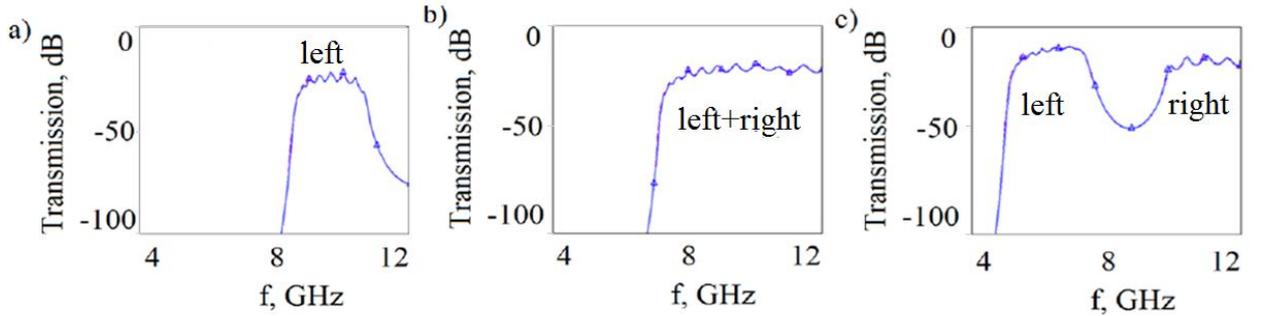

**Figure 3.** Simulated transmission magnitude $|S_{21}|$ through an array of 20 unit-cells for different values of $\varphi$: a) $\cos\varphi = 0.8$, b) $\cos\varphi = 0.4$, c) $\cos\varphi = 0.2$.

The transmission from Fig. 3 does not indicate the left-handed and right-handed behaviour directly. The scattering parameters (S-parameters) and the complex impedance of the cells can be used for clarifying whether the band is left-handed or right-handed. Both S-parameters and impedance for a magnetic fields of Fig. 3c are depicted in Fig. 4. One can see from Fig. 4a that the structure demonstrates almost full reflection ($S_{11}$=0 dB) within the frequency range of 9-11 GHz. According to Fig. 4b, the edges of the reflection band coincide with the frequencies, at which the imaginary part of the impedance curves,

Im(Z,f), change their sign. The left-handed behaviour, by definition of the model, corresponds to the reactive impedance of the parallel elements Im($Z_{shunt}$) > 0 (inductor) and, simultaneously, the reactive impedance of the series elements Im($Z_{series}$) < 0 (capacitor), while the right-handed behaviour occurs for opposite signs of these impedances. In Fig. 4c, one can see the impedance of serial and parallel elements, which corresponds to the bands shown in Fig. 3b. In this case, $f_{c1} = f_{c2}$ and the change of signs of the imaginary parts of the impedance appears at the same frequency. Since there is no band where the impedances are simultaneously both negative or positive, no rejection gap exist between the right- and left-handed frequency bands, as shows Fig. 2b. Note, however, that the absolute value of the transmission is far from 100% even within the transmission bands. This is due to the impedance mismatch between the characteristic impedance of the line and the standard ports of 50 Ω considered in the simulation.

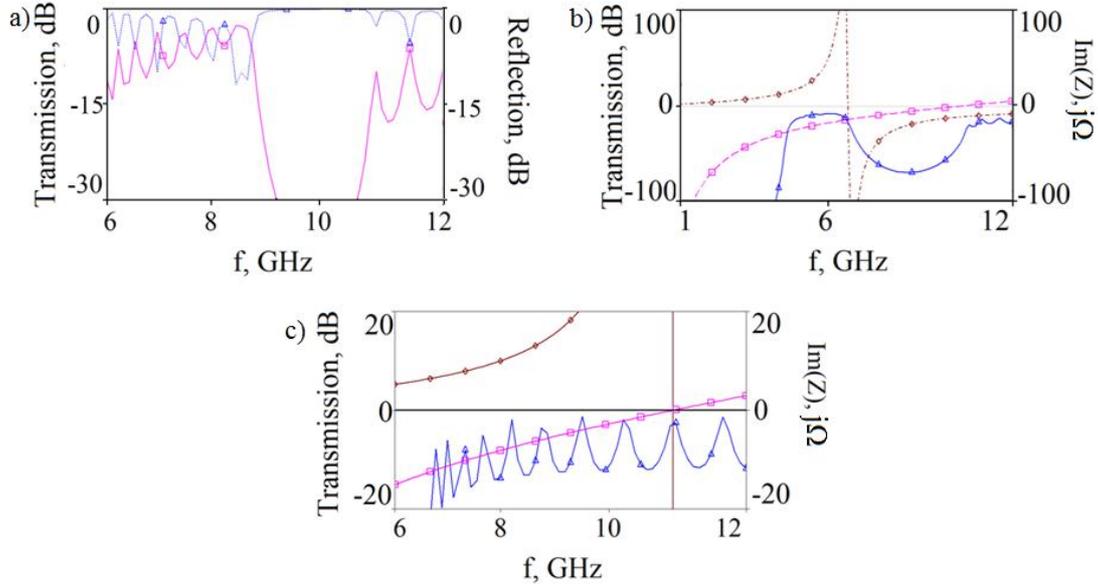

**Figure 4**. a) Simulated scattering parameters |$S_{11}$| (blue, right axis) and |$S_{21}$| (magenta, left axis) for 20 unit-cells. b)Transmission through 20 unit-cells (blue, left axis) and impedance of serial (magenta, right axis) and parallel (blue, right axis) elements for cos φ = 0.2 and $f_{c1} \neq f_{c2}$, c) Transmission through 20 unit-cells (blue, left axis) and impedance of serial (magenta, right axis) and parallel (brown, right axis) elements for cos φ = 0.4 and $f_{c1} = f_{c2}$.

## 3. Experiment.

A number of features of the experimental layout has to be restricted to the resolution of a fabrication process of the Josephson junctions. Fortunately, the available to us trilayer process Nb-AlOx-Nb ensures sufficient compactness of the unit cell. This technology process allows to use parallel plate capacitors, formed by the two niobium layers and the insulation layer from anodic oxide $Nb_2O_5$ for all capacitors. Note here, that the SQUIDs are operating in the superconducting state, and no resistive shunts are used. The layout of the cell, the optical image of three experimental cells and the wiring at the sample holder are illustrated in Fig. 5a-c. All six Josephson junctions of each unit cell, which are shown in Fig. 5a, are fabricated on the common island. The areas denoted by 5 in Fig. 5a and in dark blue in Fig. 5b are spots where the anodization wiring is etched away at the final stage of the fabrication process. A chip of the size 4x4 $mm^2$ contains two completely separated rf transmission lines (i. e. two samples). The two lines are different: one contains 10 and another contains 20 unit-cells per line. The experimental setup is described in Fig. 6a. The samples are measured using an Agilent PNA-X network analyzer in a dry close-cycle cryostat at temperatures about 2 K. Additional attenuators (40 dB) are inserted into the input line in order to reduce the external noise and make sure that the probe rf current would not exceed the critical current of the Josephson junctions. Due to the low probe current, it was necessary to boost the output signal using a low-noise amplifier (LNA). The magnetic field was applied by a superconducting coil external to the sample holder; the field magnitude is proportional to the current through the coil that is shown in the vertical axis of the following plots.

The microwave calibration is not an easy task for cryogenic measurements, as detailed in Ref. [13]. Depending on the experimental setup, a proper calibration might not be possible. This is why in this work a virtual calibration for the transmitted signal is attempted. For doing this, we averaged the response $S_{21}(f,H)$ for each frequency point, f, over the whole set of data for magnetic field, H, and subtracted the resulting relation $S_{21}^{av}(f)$ from the measured data. This procedure is feasible since we are interested in detecting the variations caused by the magnetic field at given frequency, not in frequency variation for a static magnetic field. Since we assume that the relative weight of the large deviation of *S* caused by *H* is small, the averaged response $S_{21}^{av}(f)$ should deviate slightly from the "true background". The transmission through the superconducting LHTL with 10 cells is shown in Fig. 6b. No virtual calibration used here.

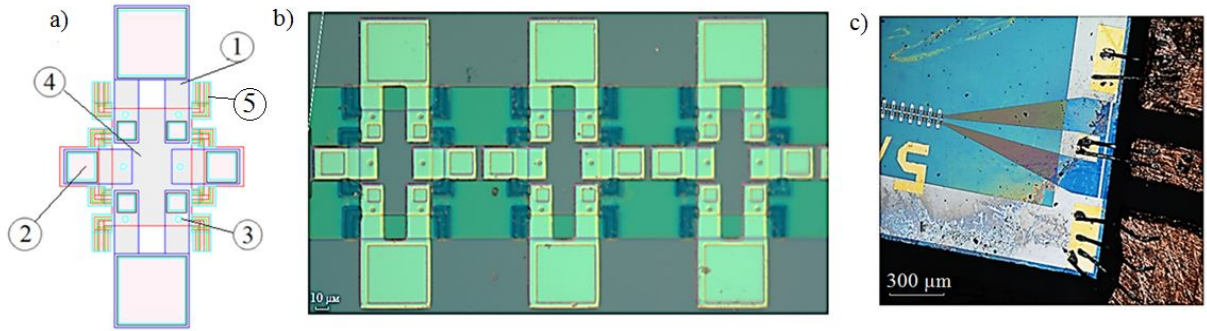

**Figure 5**. Experimental sample close-up. a) Cell layout: 1 - SQUID loop with large capacitor to ground, 2 - via bridge connecting neighboring cells, 3 - one of six Josephson junctions of the cell, 4 - bottom electrode is the central island of the cell. b) Optical microscope image of three cells of the LHTL; c) Optical microscope image of the connection to the 50-Ω CPW.

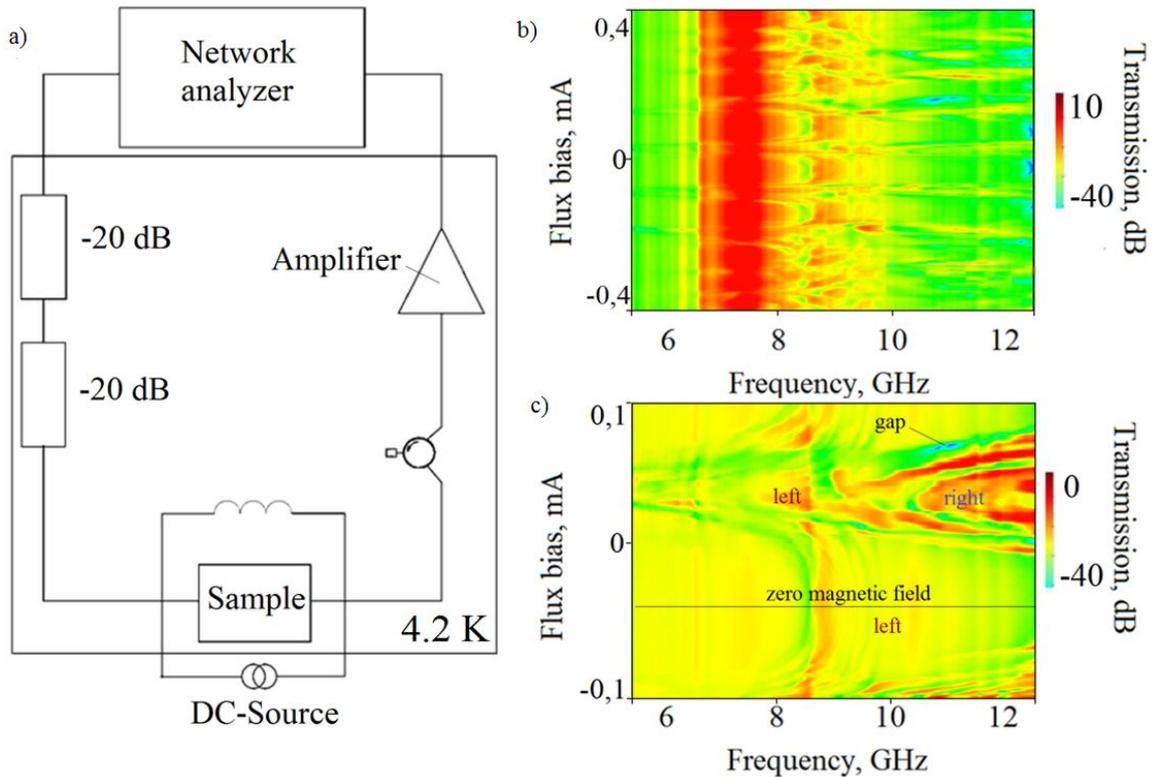

**Figure 6:** a) Scheme of the experimental setup. b) Measured transmission through the transmission line containing 10 cells as function of magnetic field. The magnetic field ranges from -0.4 mT to 1 mT. No virtual calibration is used. c) Measured transmission with subtracted background (virtual calibration) for 20-cell line as function of magnetic field. The magnetic field range is between -0.2 mT to 0.6 mT.

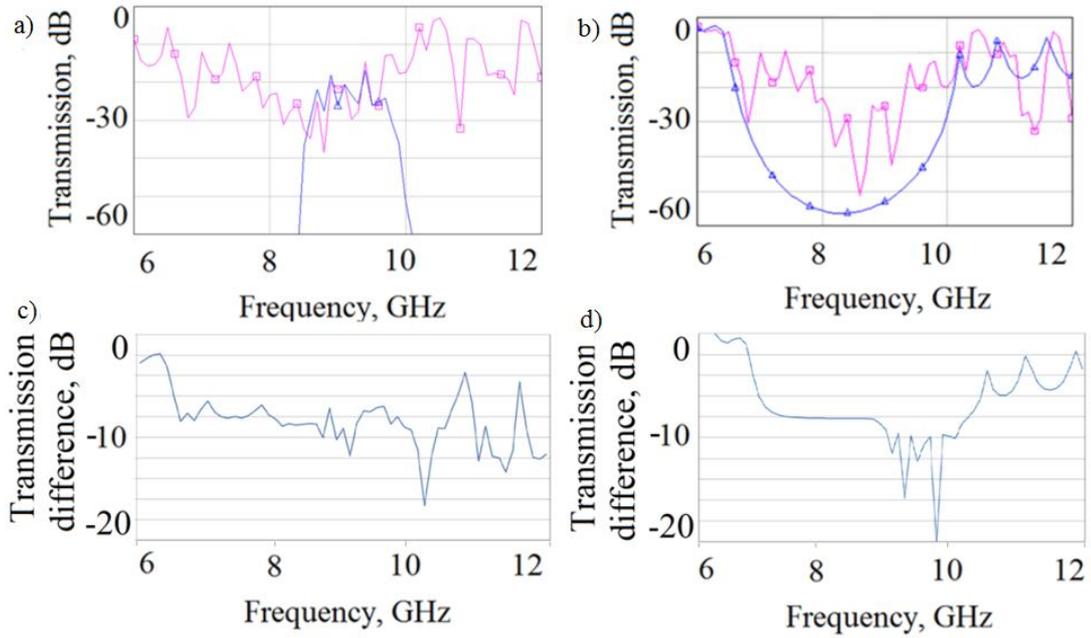

**Figure 7:** a) Simulated transmission through the line containing 20 unit cells in presence of the non-ideal (rf-leaky) sample holder (magenta) and simulation for an ideal (no-leak) sample holder (blue) for a value of cos φ = 0.8. b) Same data and color coding as in a), but with cos φ = 0.2. c) Expected response to magnetic field (from cos φ = 0.2 to cos φ = 0.8) for leaky holder as the difference between in transmission a) and b). d) The same difference as in c) but with an ideal sample holder.

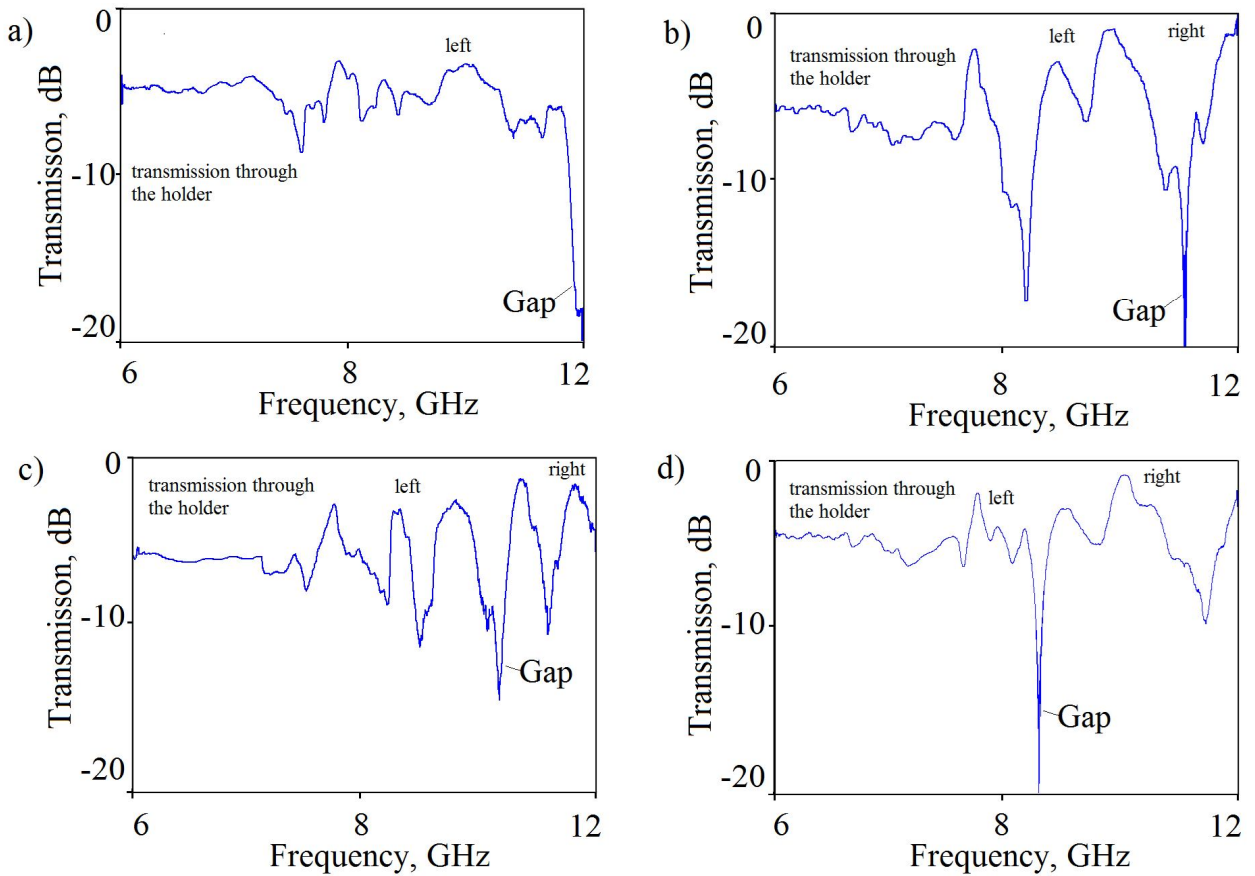

**Figure 8.** Measured transmission through 20-cell line for various magnetic fields: a) 0 mT, b) 0.15 mT, c) 0.25 mT, and d) 0.3 mT.

One can see from Fig. 6b that the transmission through the line with 10 cells changes periodically with the magnetic field, as expected due to flux quantization. In Fig. 6c, we show the transmission with subtracted background $S_{21}^{av}(f)$ for the line with 20 cells. Here, periodic changes of the transmission with the magnetic field are also observed. At zero field ($\varphi = 0$), only one transmission band is seen below 12 GHz. At larger fields $cos\varphi$ is close enough to zero, and two pass bands divided by a gap can be seen, as predicted by simulations.

We also observed some discrepancies between simulation and experiment. First, there is moderate transmission below 8 GHz, whereas the simulations predicts negligible transmission (the rejection dip). Second, the frequency range of the rejection band is narrower than predicted. Both these issues can be assigned to a parasitic stray transmission through the sample holder. We simulated this parasitic transmission. It is depicted in Fig. 7 by the pink curve. The measured rf leak is consistent with the level estimated from detailed electromagnetic simulations of our sample holder. One can see from Fig. 7a that it is hard to unambiguously extract the cut-off frequency from the experimental data at zero field, not to mention the rejection band limits at relatively large magnetic fields. The unwanted transmission through the holder appears due to, at least, two factors: i) the holder creates an additional path for the signal, ii) the characteristic impedance of the line is strongly mismatches with 50-Ohm input-output lines that is mentioned already above. Both these features lead to the significant enhancement of the unwanted, but unavoidable rf background and large standing waves in the transmission spectrum as presented in Fig. 7a,b. To clarify the effect of the rejection gap we subtracted transmission in smaller magnetic field ($cos\varphi = 0.8$), from the transmission with larger magnetic field for two cases: i) with leaky model of the sample holder (in Fig. 7c) and ii) with ideal (no-leak) sample holder (in Fig. 7d). From these data one can see that these differential signals are expected to be narrow-band in spite the wide-band feature is predicted for an ideal sample-holder. The differential signal is predicted to have few dips between the left- and right-handed bands, i.e. between approximately 7 and 11 GHz as presented in Fig. 7c, d. In this frequency range the transmission is lower in larger magnetic fields than in smaller ones. For both the leaky and the ideal holders the effect is qualitatively the same. This behavior provides strong indication of a possibility of detection the predicted gap even with the non-ideal sample holder. The experimental data demonstrate a number of narrow dips as depicted in Fig. 8. Our virtual calibration via averaging is rather qualitative, so we did not succeed in complete removal of the effect of standing waves in Fig. 8. Since the impedance of the line is changes, the reflected signal (standing wave) also changes with the magnetic field. Nevertheless, the predicted narrow gap is clearly observable: when increasing the magnetic field, it moves to lower frequency and finally disappears thus merging the left and right-handed bands. Further increase of the magnetic field leads to re-appearing of the gap and makes it deeper again as presented in Fig. 8. In spite we have observed the rather narrow-band gap, this result is supported with the predicted performance as from Fig. 7c,d.

## 4. Conclusion.

The concept of superconducting magnetic-field tunable left-handed transmission line with Josephson junctions has passed experimental verification within the frequency range of 6-12 GHz. The calculated dispersion of the line shows left-handed behaviour within the lower frequency transmission band. Experimental data demonstrate tunability of both the transmission and rejection bands by applying a dc magnetic field. The magnetic field allows us to shift the rejection band between the left- and right-handed transmission. Moreover, for a limited frequency range near 10 GHz, the transmission line can be transformed from left-handed to right-handed by changing the magnetic field. The presented experiments are noticeably obscured by parasitic transmission through the sample holder. An improved design of the sample holder is currently under development and we expect to obtain more detailed data in the nearest future. A good potential for improvement can be found on the way of designing a better impedance match between the sample and the standard 50-Ohm cables.

**Acknowledgement.**
This work was supported in part by the Ministry of Education and Science of the Russian Federation, by the Russian Foundation of Basic Research, and by the Deutsche Forschungsgemeinschaft (DFG) and the State of Baden-Württemberg through the DFG Center for Functional Nanostructures (CFN).

**References**
[1] Veselago V G 1967 The Electrodynamics of Substances with Simultaneously Negative Values of ε and μ *Usp. Fiz. Nauk* vol 92 h 517-526
[2] Shen J Q 2008 Introduction to the Theory of Left-Handed Media *arXiv cond-mat* 0402213
[3] Pendry J, Smith D 2004 Reversing Light with Negative Refraction *Physics Today* vol 57 no 6 p 37–43

[4] Caloz C, Itoh T 2003 Novel microwave devices and structures based on the transmission line approach of meta-materials *IEEE MTT-S Int. MTT Symp. Dig.* Vol. 1 pp. 195-198

[5] Eleftheriades V, Klyer A, Kremer P 2002 Planar negative refractive index media using periodically l-c loaded transmission lines *IEEE Trans. Microw. Theory Tech.* vol 50 no 12 pp 2702-2712

[6] Wang Y and Lancaster M J 2006 High-temperature superconducting coplanar left-handed transmission lines and resonators *IEEE Trans. Appl. Supercond.* 16 1893

[7] Salehi H, Majedi A and Mansour R 2005 Analysis and design of superconducting left-handed transmission lines *IEEE Trans. Appl. Supercond.* 15 pp 996-999

[8] Durán-Sindreu M, Damm Ch, Sazegar M, Zheng Yu, Bonache J, Jakoby R, Martín F 2011 Electrically Tunable Composite Right/Left Handed Transmission-Line based on Open Resonators and Barium-Stronium-Titanate Thick Films *Microwave Symposium Digest (MTT), IEEE MTT-S International*

[9] Hutter C, Tholen E A, Stannigel K, Lidmar J, Haviland D B 2011 Josephson junction transmission lines as tunable artificial crystals *Phys. Rev. B* 83 014511

[10] Jung P, Butz S, Shitov S V and Ustinov A V 2013 Low-loss tunable metamaterials using superconducting circuits with Josephson junctions *Appl. Phys. Lett.* 102 062601

[11] Schmidt V V *The physics of superconductors* Eds. P Mueller and A V Ustinov (Berlin: Springer, 1997)

[12] Pozar D 1998 *Microwave engineering* (Toronto: John Wiley & Sons) 2nd edition pp 424-427, 162

[13] Anlage S M, Yeh J-H 2013 In situ broadband cryogenic calibration for two-port superconducting microwave resonators *Rev. Sci. Instrum.* 84 034706